%% file: sec_000_main.tex
\documentclass[letterpaper, 10 pt, conference]{ieeeconf}  %

\IEEEoverridecommandlockouts                              %

\usepackage{balance} %

\usepackage{algorithm,algorithmic}
\usepackage{amsmath}
\usepackage{amsfonts}
\usepackage{color}
\usepackage{mathtools}

\newtheorem{Proposition}{Proposition}[section]
\DeclareMathOperator*{\argmax}{argmax}
\DeclareMathOperator*{\argmin}{argmin}

\definecolor{darkgreen}{rgb}{0,0.5,0}

\title{\LARGE \bf Robust Electric Vehicle Balancing of Autonomous Mobility-on-Demand System: A Multi-Agent Reinforcement Learning Approach}

\author{
{Sihong He$^{1}$}  \and {Shuo Han$^{2}$} \and {Fei Miao$^{1}$} 
\thanks{$^{1}$Sihong~He and Fei~Miao are with the Department of Computer Science and Engineering, University of Connecticut, Storrs, CT. {\tt\small \{sihong.he, fei.miao\}@uconn.edu}. }
\thanks{$^{2}$Shuo Han is with the Department of Electrical and Computer Engineering, University of Illinois, Chicago. {\tt\small hanshuo@uic.edu}. }
\thanks{This work is supported by National Science Foundation under Grants CNS-1952096, CMMI-1932250, and CNS-2047354.}
}

\newcommand{\BibTeX}{\rm B\kern-.05em{\sc i\kern-.025em b}\kern-.08em\TeX}

\begin{document}

\maketitle
\thispagestyle{plain}
\pagestyle{plain}

\begin{abstract}
Electric autonomous vehicles (EAVs) are getting attention in future autonomous mobility-on-demand (AMoD) systems due to their economic and societal benefits. However, EAVs' unique charging patterns (long charging time, high charging frequency, unpredictable charging behaviors, etc.) make it challenging to accurately predict the EAVs supply in E-AMoD systems. Furthermore, the mobility demand's prediction uncertainty makes it an urgent and challenging task to design an integrated vehicle balancing solution under supply and demand uncertainties. Despite the success of reinforcement learning-based E-AMoD balancing algorithms, state uncertainties under the EV supply or mobility demand remain unexplored. In this work, we design a multi-agent reinforcement learning (MARL)-based framework for EAVs balancing in E-AMoD systems, with adversarial agents to model both the EAVs supply and mobility demand uncertainties that may undermine the vehicle balancing solutions. We then propose a \textbf{r}obust \textbf{E}-AMoD \textbf{Ba}lancing \textbf{MA}RL (REBAMA) algorithm to train a robust EAVs balancing policy to balance both the supply-demand ratio and charging utilization rate across the whole city. Experiments show that our proposed robust method performs better compared with a non-robust MARL method that does not consider state uncertainties; it improves the reward, charging utilization fairness, and supply-demand fairness by 19.28\%, 28.18\%, and 3.97\%, respectively. Compared with a robust optimization-based method, the proposed MARL algorithm can improve the reward, charging utilization fairness, and supply-demand fairness by 8.21\%, 8.29\%, and 9.42\%, respectively.
\end{abstract}

\section{Introduction}
\input{sec_001_intro}

\section{Related Work}
\label{sec:related_work}
\input{sec_002_related_work}

\section{Robust Multi-Agent Reinforcement Learning Framework for E-AMoD Balancing}
\label{sec:formulation}
\input{sec_003_formulation}

\section{Algorithm}
\label{sec: algrithm}
\input{sec_004_algorithm}

\section{Experiment}
\label{sec: experiment}
\input{sec_005_exp}

\section{Conclusion}
\label{sec:conclusion}
Electric autonomous vehicles (EAVs) are playing important roles in future autonomous mobility-on-demand (AMoD) systems. However, it remains challenging to address E-AMoD system's state uncertainties caused by EAVs' unique charging patterns and AMoD systems' mobility demand in algorithm design. In this work, we design a robust MARL framework to balance mobility supply-demand ratio and charging utilization rate for E-AMoD systems under EAV supply and mobility demand uncertainties. The mobility demand and EAV supply uncertainties are captured by adversarial agents that can cause state uncertainties during the training process. We then design a REMABA algorithm with Dykstra's projection and policy regression. Experiments show that our proposed robust algorithm performs better in terms of reward, increases charging utilization fairness and supply-demand fairness by 19.28\%, 28.18\%, and 3.97\%, respectively, compared with non-robust MARL-based methods that ignore system uncertainties. Compared with a robust optimization-based method, the proposed MARL algorithm can increase reward, charging utilization fairness and supply-demand fairness by 8.21\%, 8.29\%, and 9.42\%, respectively.

\bibliographystyle{ieeetr} 
\bibliography{sample}

\end{document}

%% file: sec_001_intro.tex
The Electric Autonomous Mobility on Demand (E-AMoD) system is an energy-efficient and sustainable alternative to private urban mobility by using Electric Autonomous Vehicles (EAVs) to provide one-way rides to passengers \cite{zardini2021analysis_survey}. With E-AMoD, passengers express their travel needs through mobile applications, phone reservations, or street hails. Vacant EAVs then provide passengers with ride services \cite{shaheen2020mobility}. Besides, the concept of shared use of a vehicle reduces the total travel costs and urban infrastructure needed for parking and increases vehicle utilization \cite{wen2017rebalancing}. In light of it, Pony.ai piloted its first commercial E-AMoD service in 2019~\cite{pony_ai}. E-AMoD system has therefore been recognized as one of the most promising systems to address the challenge of the growing mobility needs and oil dependency. %

However, E-AMoD systems operation presents challenges in multiple aspects. Passenger demands are usually asymmetry distributed because of the spatiotemporal urban mobility nature \cite{gammelli2021graph_cdc}. This problem gets severe in rush hours when passengers travel in similar directions, such as from residential areas to work areas at morning peak. Without re-balancing, many idle EAVs can potentially aggregate in particular areas.

\begin{figure}
	\centering
	\includegraphics [width=0.85\linewidth]{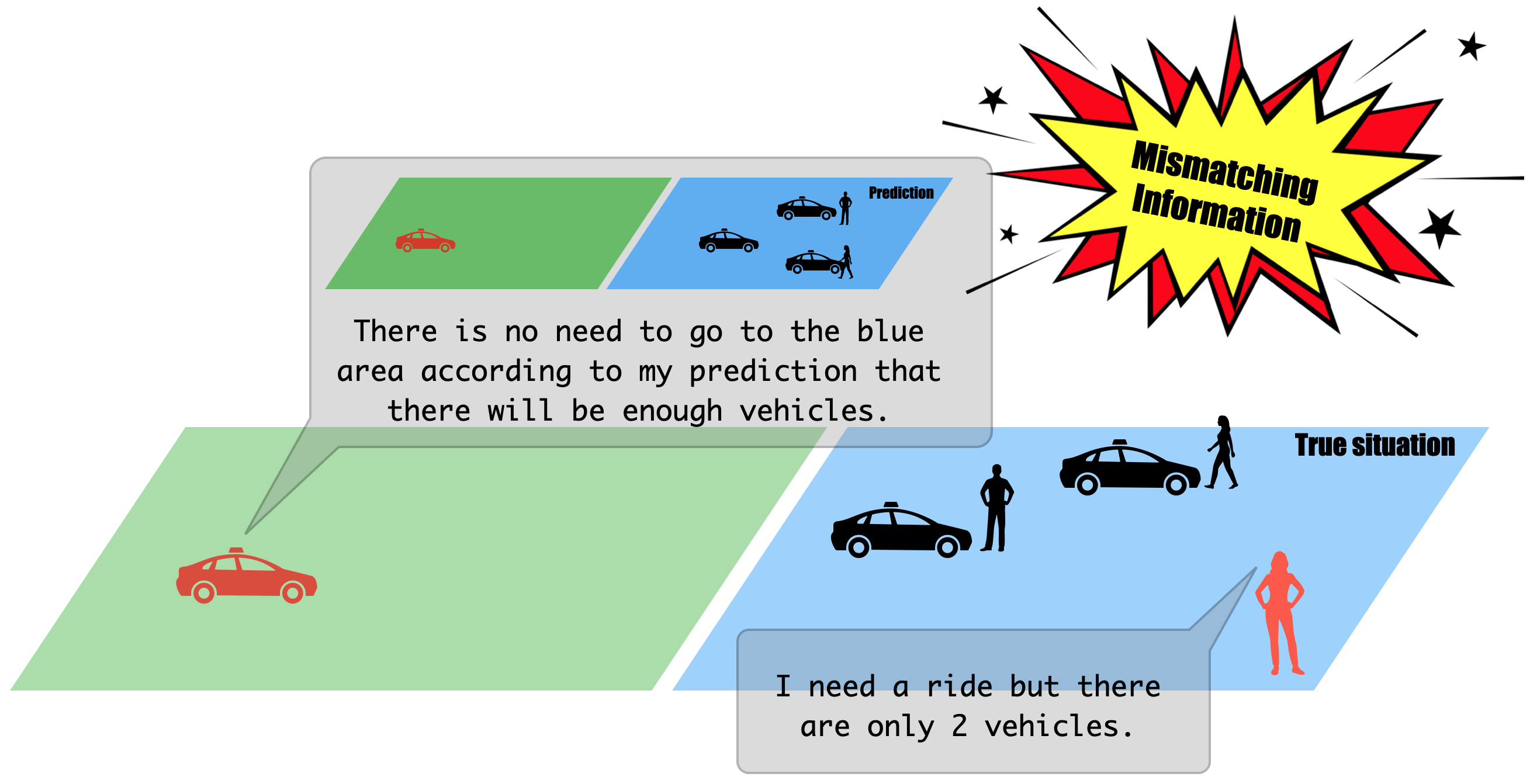}
	\vspace{-10pt}
	\caption{State uncertainties (the inaccurate information about mobility demand and vehicle supply) degrade the performance of vehicle balancing methods. The decision maker gets state information that there will be 3 vehicles and 2 passenger demands in the blue area. So the vehicle balancing decision is to keep the red vehicle in the green area to avoid energy consumption. But the true situation is that there are 2 vehicles and 3 passengers in the blue area which means the red vehicle should move to the blue area.}
	\label{fig_mismatch}
 	\vspace{-10pt}
\end{figure}

Therefore, we study the vehicle balancing problem \cite{zardini2021analysis_survey, dro_he} for E-AMoD systems in this work. In specific, we consider two main scheduling tasks for operating an EAV fleet, namely (i) vehicle rebalancing, to reposition idle EAVs to other locations, and (ii) charging scheduling, i.e., assigning the charging stations for low-battery EAVs. Recent work regarding vehicle balancing methods can be classified into three typical categories: (1) heuristic methods; (2) optimization-based approaches and (3) Reinforcement Learning (RL)-based approaches. The rule-based heuristics usually lead to sub-optimal solutions \cite{liu2019dynamic_heuristic, vandael2015reinforcement_heuristic}. Optimization-based approaches usually propose an optimization problem based on the system dynamic model \cite{AMoD_queue, xie2019optimal, ddrobust_Miao}. The performance of these methods is therefore heavily affected by modeling knowledge. RL-based methods formulate the problem as a Markov Decision Process (MDP) and apply RL algorithms to find the optimal balancing policy \cite{EVcharge_19, he2020spatio}. We provide more discussions in the Related Work section.

Nevertheless, state uncertainties from different sources may result in the degradation of vehicle balancing decisions. We provide an instance in Fig. \ref{fig_mismatch}, when there is a difference between the true system information and the prediction of mobility demand and vehicle supply, the vehicle balancing methods do not have performance guarantees. It is necessary to consider multiple uncertainties in E-AMoD systems. The EAVs' unique charging patterns (long charging time,  high charging frequency and unpredictable charging behaviors, etc.) increase systemic supply uncertainties~\cite{pricing_20, p2charge, jointcharging}. E-AMod systems' characteristics (unpredictable long daily driving time, uncertain sporadic demands and dispersed mobility patterns, etc.) also increase the uncertainties in demands prediction~\cite{epat, dddro_tcps20}. Existing EAMod balancing algorithms usually do not consider EAV supply uncertainties~\cite{p2charge, pricing_20}, or only consider mobility demand uncertainties~\cite{mpcmod_icra16, dddro_tcps20}.

Hence, in this work, we develop a multi-agent reinforcement learning (MARL) based robust EAV balancing framework for E-AMoD systems, in which region agents make fair vehicle balancing decisions, and adversarial agents  model the state uncertainties including both passenger demand and EV supply uncertainties. Our \textit{main contributions} are as follows:

\begin{itemize}
    \item To the best of our knowledge, we are the first to formulate the E-AMoD system vehicle balancing problem under demand and supply uncertainties as a robust multi-agent reinforcement learning problem under state uncertainties. Via a proper design of the agent, state, action, and reward, we set the goal of the problem as balancing the whole city's charging utilization and mobility service quality.
    \item We design a \textbf{r}obust \textbf{E}-AMoD \textbf{Ba}lancing \textbf{MA}RL algorithm (REBAMA) to train robust policies for providing fair mobility and charging services. It adopts the centralized training and decentralized execution framework with Dykstra’s projection and policy regression to keep actions from violating real-world constraints during policy updates.
    \item We run experiments based on real-world E-taxi system data. Experiments show that our proposed REBAMA algorithm performs better in terms of reward, charging utilization fairness and supply-demand fairness, which are increased by 19.28\%, 28.18\%, and 3.97\%, respectively, compared with a non-robust MARL-based method that does not consider system uncertainties.
\end{itemize}

%% file: sec_002_related_work.tex
Researchers have studied EV charging scheduling, AMoD system vehicle rebalancing, and joint scheduling using rule-based heuristic, optimization-based, or reinforcement learning-based approaches. For instance, Liu et al.~\cite{liu2019dynamic_heuristic} and Vandael et al.~\cite{vandael2015reinforcement_heuristic} proposed heuristic schemes of vehicle rebalancing and individual EV charging, respectively. These heuristic methods usually lead to sub-optimal solutions~\cite{zardini2021analysis_survey}. 

Optimization-based methods first design an optimization problem based on Model Predictive Control (MDP), then solve it at each time step to yield a sequence of balancing actions over a receding horizon. But only the first balancing action is executed. Under this category, charging scheduling approaches consider different objectives  have been proposed, such as reducing charging delays or balancing the charging tasks in charging stations with limited resources~\cite{yan2018employing, bCharge, EVAMoD_tcns19}, reducing idle distance or idle time~\cite{yan2018employing, p2charge}, and improving drivers revenue~\cite{xie2019optimal}, etc. AMoD systems' vehicle rebalancing approaches with various objectives have been designed, such as improving service quality~\cite{predictmod_icra17, Morari_rideshare}, maximizing the number of served passengers with a reduced number of vehicles~\cite{mpcmod_icra16, mod_iros18, DDmpcmod_icra18}. These optimization-based approaches usually rely on precise modeling of the complex probability state transition model of E-AMoD systems and future mobility demand and EAV supply predictions. Therefore, they are sensitive to model uncertainties, prediction errors and measurement inaccuracy.

RL-based methods formulate the vehicle balancing problem as a Markov Decision Process and apply RL algorithms to find the optimal balancing policy. Wen et al. \cite{wen2017rebalancing} applied Deep Q-Network (DQN) to study the vehicle balancing problem. Holler et al. developed an Actor-Critic-based fleet management algorithm to reposition vehicles \cite{holler2019deep_icdm, gueriau2018samod}. Various RL algorithms \cite{EVcharge_19} such as contextual DQN and A2C \cite{lin2018efficient, gueriau2018samod}, spatio-temporal capsule-based Q-learning \cite{he2020spatio}, mean-field multi-agent RL \cite{jointcharging} algorithms have been proposed to solve the vehicle balancing problem. Compared to optimization-based vehicle balancing methods, RL-based methods can handle a larger-scale problem in practice by incorporating with function approximation scheme, and relax the dependence on the modeling of E-AMoD systems' complex dynamics~\cite{he2022robust_icra}. However, passenger mobility demand or EV supply uncertainties are not considered in RL methods yet. For the first time, we consider passenger mobility demand or EV supply uncertainties as system state information uncertainties, and propose robust MARL-based problem formulation and algorithm for E-AMoD system balancing under state uncertainties.

%% file: sec_003_formulation.tex
\subsection{Problem Statement}
We consider the problem of managing a large-scale EAVs fleet to provide fair and robust E-AMoD service. The proposed method should (i) rebalance idle EAVs for providing fair ride service on the passenger's side; (ii) allocate low-battery EAVs to charging stations for balanced charging service on the EAVs' side; (iii) be robust to mobility demand and EAV supply uncertainties. 
We consider that the city is divided into $N$ regions according to a pre-defined partition method~\cite{ddrobust_Miao, dro_he}, and a day is divided into equal-length time intervals. At each time interval $[t, t+1)$, passengers' ride requests and low-battery EAV's charging needs emerge in each region. After the locations and status of each EAV are observed and updated, a local controller assigns available EAVs to pick up existing passengers in the request queue according to specific trip assignment algorithms, such as methods designed in the literature~\cite{survey_19}, and assigns EAVs that need to be charged to charging stations~\cite{wang2019shared}. Then the predicted passenger demand and available charging spots at each region for time interval $[t, t+1)$ are updated, and a system-level EAV balancing decision is calculated according to the algorithm designed in this work. 

Some assumptions about the model and algorithm are considered as follows. We consider four dynamic statuses for one EV: {\em vacant}, {\em occupied}, {\em low-battery} and {\em still}, similar to the literature~\cite{dro_he}. {\em Vacant} means the EV is not serving any passengers, i.e. idle and its remaining battery exceeds a threshold. The algorithm dispatches vacant EAVs to stay in the current region to pick up passengers or to move to other regions according to the future predicted passenger mobility demand in the following time interval. When a vacant EV picks up passengers, it turns to {\em occupied} status and we have no dispatch command for it until it becomes vacant again. An occupied EV becomes vacant again when it drops off passengers and still has enough battery. Once a vacant EV's battery level is lower than a threshold, it becomes a {\em low-battery} EV. The algorithm will assign low-battery EAVs to some regions for charging according to the availability of charging stations. When low-battery EAVs enter charging stations, they become {\em still} EAVs. Still EAVs become vacant when they finish charging and leave the charging stations.

We focus on a computationally tractable system-level EV balancing algorithm design such that both passenger demand and EV supply uncertainties are considered to maximize the expected total reward of the system or the entire city. The local trip and charging assignment algorithm are out of the scope of this work. For notation convenience, the parameters and variables definition in the following parts of this section omit the time index $t$ when there is no confusion.

\subsection{Zero-Sum Stochastic Game}
\label{sec_game}
We formulate the vehicle balancing problem for E-AMoD systems as a zero-sum stochastic game $\mathcal{G}$ between a set of region agents $\mathcal{N}_r$ and a set of adversarial agents $\mathcal{N}_a$. The \textit{region agent} is designed for each region to make dispatch decisions for vacant and low-battery EAVs at every time step. This distributed agent setting is more reasonable for large-scale fleet management than a single agent setting because the action space can be prohibitively large if we use a single agent~\cite{ev_kdd19}.  The \textit{adversarial agent} is designed to model uncertainties that may be caused by delayed information, missing data, inaccurate measurement and prediction errors. The adversarial agent can make dispatching decisions robust to modeling errors and real-world uncertainties by altering the protagonist's observation~\cite{pattanaik2017robust} in the training process. We define $\mathcal{S}_i$ as the state space for region $i$ shared by both the $i$-th region agent $i \in \mathcal{N}_r$ and adversarial agent $i \in \mathcal{N}_a$, and $\mathcal{S}\coloneqq \mathcal{S}_{1} \times \dots \times \mathcal{S}_{N}$ as the joint state space, then $\mathcal{G} \coloneqq  (\mathcal{N}_a, \mathcal{N}_r, \mathcal{S}, \mathcal{A}_a, \mathcal{A}_r, r, T)$. We define $\mathcal{A}_{a,i}$ and $\mathcal{A}_{r,i}$ as the action space of the adversarial agent and region agent of region $i$, and $\mathcal{A}_a \coloneqq \mathcal{A}_{a,1} \times \dots \times \mathcal{A}_{a,N}, \mathcal{A}_r \coloneqq \mathcal{A}_{r,1} \times \dots \times \mathcal{A}_{r,N}$ as the joint action space of adversarial and region agents, respectively. The map $r: \mathcal{S} \times \mathcal{A}_r \times \mathcal{A}_a \rightarrow \mathbf{R}$ is the reward function shared by all agents. The state transition probability function is $T: \mathcal{S}\times \mathcal{A}_r \times \mathcal{A}_a \rightarrow \Delta(\mathcal{S})$, where $\Delta(\mathcal{S})$ represents the set of probability distributions over the joint state space $\mathcal{S}$. $T (s'|s,a_r,a_a)$ is the probability of next state $s' \in \mathcal{S}$ given the current state $s$ and the adversarial joint actions $a_a$ and balancing decisions $a_r$ (region agents' joint actions). We formally define the states and actions in the next section. At each time step $t$, for all $i \in \mathcal{N}_a$, adversarial agent $i$ observes the true state information and chooses its action $a_{a,i}^t$ according to a policy $\pi_{a,i}: \mathcal{S}_i \rightarrow \Delta(\mathcal{A}_{a,i})$ to manipulate $s_i$. Region agent $i$ can only observe the perturbed state information and chooses its action $a_{r,i}^t$ according to a policy $\pi_{r,i}: \mathcal{S}_i \rightarrow \Delta(\mathcal{A}_{r,i})$. We define the adversarial agents' joint policy $\pi_a = \prod_{i \in \mathcal{N}_a}\pi^{a,i}: \mathcal{S} \rightarrow \Delta(\mathcal{A}_a)$ and region agents' joint policy $\pi_r = \prod_{i \in \mathcal{N}_r}\pi^{r,i}: \mathcal{S} \rightarrow \Delta(\mathcal{A}_r)$. After all region agents execute their actions, they get a shared reward $r^t$ and all adversarial agents get a shared opposite reward $-r^t$. The value functions are defined as the discounted return of region agents, i.e., 
action value function $q^{\pi_r, \pi_a}(s,a_r,a_a) = \mathbb{E}[ \sum_{t=1}^{\infty} \gamma^{t-1} r^t | s^1 = s, a_r^1 = a_r,a_a^1 = a_a,a_r^t \sim \pi_r(\cdot | \tilde{s}^t), a_a^t \sim \pi_a(\cdot | s^t)]$, $\tilde{s}^t = f(s^t, a_{a}^t)$, where $f: \mathcal{S}_i \times \mathcal{A}_{a,i} \rightarrow \mathcal{S}_i$ is the state perturbation function which describes the relationship between perturbed state, true state and adversarial agents' actions. We use $\tilde{s}^t = f(s^t,a_a^t)$ to denote $\tilde{s} = (\tilde{s}_1^t, \cdots, \tilde{s}_N^t)$ where $\tilde{s}_i^t = f(s_i^t, a_{a,i}^t)$, $\forall i = 1, \cdots, N$. The power of adversaries can be restricted by adding constraints to the perturbed state such as $\tilde{s} \in B(s, \epsilon)$ where $B(s, \epsilon)$ is a $\epsilon$-ball centered in the true state $s$, or by carefully defining the state perturbation function. Thus, we trade between the robustness and performance of agent policies by adjusting adversaries' perturbation power \cite{he2023robust, han2022solution}. In equation \eqref{def:map}, we give the formal definition of the perturbation function $f$ in our vehicle balancing problem of E-AMoD systems. Our goal is to solve the following min-max problem to get a robust region agents' joint policy.
\begin{align}
\label{minmax}
    \max_{\pi_r}\min_{\pi_a}\mathbb{E}[v^{\pi_r, \pi_a}(s)]
\end{align}
Zero-sum games involve two players/teams with conflicting objectives, where one side's gain is the other side's loss. This adversarial nature makes them suitable for modeling robustness problems since they capture situations where a system needs to perform well under strong perturbations or uncertain conditions. While the zero-sum game model does not explicitly capture road capacity and speed limits, it can still incorporate these aspects indirectly through the reward design and action spaces.

\subsection{MARL Problem Formulation}
In this section, we formally define the state, action, reward and policy in our robust MARL framework.

\paragraph{State} A state $s_{i}^t \in \mathcal{S}_i$ of a region $i$ contains a vector that indicates its spatiotemporal status from both the local view and global view of the city at  time $t$ (we omit the time subscript $t$ later for convenience), and $s\coloneqq (s_1, \dots,  s_N)$. We define the state $s_i=\{s^{local}_i, s^{global}_i\}$, where $s^{local}_i = (V_i, L_i, d_i, ST_i, ES_i, SP_i)$ as the state from the local view, denoting the predictions about the amount of vacant EAVs, low-battery EAVs, mobility demand, still EAVs, empty charging spots, and total charging spots in region $i$, respectively. We define $s^{global}_i = (t, pos_i)$ as the spatial-temporal information from a global perspective, where $t$ is the time step index, $pos_i$ is region location information (longitudes, latitudes, boundaries, region index).

\paragraph{Region Agents Action} We use $\mathcal{N}(i)$ to denote the set of neighboring regions of region $i$, i.e., the adjacent regions according to the graph structure of the city and $n_i=|\mathcal{N}(i)| + 1$. A region agent $i$'s action $a_{r,i} = (\overline{p}, \overline{q}) \in \Delta^{n_i} \times \Delta^{n_i}$ where $\Delta^{n_i}$ denotes the probability simplex in $n_i$ dimensions. The action consists of balancing decisions for vacant EAVs to potential passenger demand $\overline{p} \in \Delta^{n_i}$, and low-battery EAVs without charging yet to potential available charging stations $\overline{q} \in \Delta^{n_i}$. Here, $p_{j \in \{1,2,...,n_i\}}$ is the $j$-th element of $\overline{p}$ that represents the percentage of vacant EAVs will move to the $j$-th region of $\{ \mathcal{N}(i) \cup \ {\text{region}\ i} \}$. For instance, $p_1= 0.1$ means $10$\% vacant EAVs are arbitrarily chosen and dispatched to the first adjacent region of region $i$. And $\overline{q}$ has a similar definition for low-battery EAVs dispatching decision. After executing region actions, the number of vacant, low-battery and still EAVs in different regions will be changed. Since the sum of the percentages of EAVs dispatched to different directions should be $1$, the dispatching actions have constraints that: $\sum_j \bar p_j = \sum_j \bar q_j \equiv 1$. 

\paragraph{Adversarial Agents Action} Adversarial agent $i \in \mathcal{N}_a$ alters the corresponding region agent's knowledge of the state by adding perturbation $a_{a,i} \in \mathcal{A}_{r,i}$. Here, $a_{a,i} = [\delta_d,\delta_c, \delta_v$] represents the volatility of predicted demand, empty charging spots, and vacant EAVs, respectively. The value of the perturbation simulates the state uncertainties from historical data;
for instance, we use box constraints that $d_l \leq \delta_d \leq d_u; c_l \leq \delta_c \leq c_u; v_l \leq \delta_v \leq v_u$ where the upper and lower bounds are determined by empirical experiments~\cite{dddro_tcps20, ddrobust_Miao}. We will also consider other more complicated formats of action space (such as second-order cone or ellipsoid) in future work. After formally defining the perturbation function $\tilde{s} = (\tilde{s}_1, \cdots, \tilde{s}_N)$, we give the formal definition of the state perturbation function $f: \mathcal{S}_i \times \mathcal{A}_{a,i} \rightarrow \mathcal{S}_i$ in our vehicle balancing problem of the E-AMoD systems. 
The altered local state $\Tilde{s}_i^{local} = (\widetilde{V}_i, \widetilde{L}_i, \widetilde{d}_i, \widetilde{ST}_i, \widetilde{ES}_i, \widetilde{SP}_i)$ 
is related to $s^{local}_i = (V_i, L_i, d_i, ST_i, ES_i, SP_i)$ and adversarial action $a_{a,i}$ by the following equation~\eqref{def:map}:
\begin{equation}
\begin{split}
    \widetilde{V}_i&= V_i + (SP_i - ES_i)\delta_c + V_i\delta_v;\\
    \widetilde{d}_i &= d_i(1+\delta_d);\\
    \widetilde{ST}_i &= ST_i - (SP_i - ES_i)\delta_c; \\
    \widetilde{ES}_i &= (SP_i - ST_i) \times I_{\{\widetilde{ST}_i < SP_i\}}.
\end{split}
\label{def:map}
\end{equation}
We denote $\tilde{s}_i = \{\tilde{s}_i^{local}, s_i^{global}\}$ which means we have $\tilde{s}_i = f(s_i = \{s_i^{local}, s_i^{global}\}, a_{a,i} = [\delta_d,\delta_c, \delta_v])$. Region agents choose their actions according to their policies and the perturbed states. After all region agents execute actions, the  system states change (e.g. number of EAVs).

\paragraph{Reward} Our goal is to optimize the system-level benefit, i.e., balanced charging utilization and fair service, hence, all region agents share common interests and the same reward function. By maximizing the shared reward, region agents are cooperating for the same goal. We let adversarial agents' reward function be the negative one of region agents. Thus, all adversarial agents aim to minimize the region agents' reward. 

We consider both the supply-demand ratio of vacant EAVs to the total mobility demand~\cite{dddro_tcps20, Morari_rideshare}, and the charging utilization rate~\cite{he2023data, p2charge, pricing_20} as service quality metrics for E-AMoD systems. A higher supply-demand ratio means a shorter waiting time for customers in one region. However, with limited EAV supply, achieving high supply-demand ratios in all regions is impossible. Keeping the supply-demand ratio of each region at a similar level allows passengers in the city to receive fair service~\cite{AMoD_queue, mpcmod_icra16}. Similarly, given the limited amounts of charging stations and spots,  to improve charging service quality and charging efficiency with limited infrastructure, balancing the charging utilization rate of all regions across the entire city is usually one objective for EV charging~\cite{EVcharge_19, p2charge}. Therefore, we define the fairness metric of charging utilization rate $u_c(s,a_r,a_a)$ and supply-demand ratio $u_s(s,a_r,a_a)$ as the negative sum of the difference between the local utilization rate (local supply-demand ratio) and the global charging utilization rate (global supply-demand ratio), respectively:
\begin{equation}
\begin{split}
    u_c(s,a_r,a_a) =   \sum^N_{i = 1} -\left\lvert  \frac{ES_i}{ST_i}- \frac{\sum^N_{j = 1}ES_j}{\sum^N_{j = 1}ST_j} \right\rvert,\\
    u_s(s,a_r,a_a)=  \sum^N_{i = 1} -\left\lvert  \frac{d_i}{V_i}- \frac{\sum^N_{j = 1}d_j}{\sum^N_{j = 1}V_j} \right\rvert,
    \label{def:u}
\end{split}
\end{equation}
A larger $u_c(s,a_r,a_a)$ or $u_s(s,a_r,a_a)$ value means a better balanced and fair charging utilization rate or supply-demand ratio among the city. Then we define the reward function $r(s,a_r,a_a)$ as a weighted sum of the city's charging utilization fairness $u_c$ and supply-demand fairness $u_s$, i.e. 
\begin{equation}
    r(s,a_r,a_a) \coloneqq u_c(s, a_r, a_a) + \beta u_s(s, a_r, a_a),
    \label{def:reward}
\end{equation} 
where $\beta$ is a positive weighted coefficient, $s, a_r, a_a$ is the joint state, joint action of all region agents, joint action of all adversarial agents, respectively.  
$r(s,a_r,a_a)$ is calculated after EAVs balance decision execution with a complex dynamic state transition process in the simulator.
One advantage of the proposed MARL method is the form of the reward function we would like to optimize does not need to satisfy the constraints as those in robust or distributionally robust optimization methods~\cite{ddrobust_Miao, dddro_tcps20}. For instance, the objective function does not need to be convex of the decision variable or concave of the uncertain parameters. %

\paragraph{Policy} All region agents share the same action space and deterministic policy, defined as $ \mu_r(\tilde{s}_i,s_{\mathcal{N}(i)}|\theta^r)$, parameterized by $\theta^r$. 
Similarly, all adversarial agents share one policy function, defined as $\mu_a (s_i,s_{\mathcal{N}(i)}|\theta^a)$, 
parameterized by $\theta^a$. We use $o_{a,i}, o_{r,i}$ to denote adversarial agent $i$ and region agent $i$' policy input, respectively. We use $\mathcal{O}$ to denote the set of all $o_{r,i},o_{a,i}$. The policy for each region relates to both its own state and its neighbors' states, since the actions of the region $i$ will affect the number of EAVs at different statuses in the neighbor regions $\mathcal{N}(i)$, and then affect the reward of the neighbor regions. This shared policy design provides a much more efficient learning procedure than training an individual policy function for each individual agent. Since the state of each agent contains spatiotemporal information, the policy of region agents is supposed to make a dispatch decision for each region according to its spatial correlation with other regions for each time step.

%% file: sec_004_algorithm.tex
\begin{figure}
	\centering
    \vspace{8pt}
	\includegraphics[width=.5\textwidth]{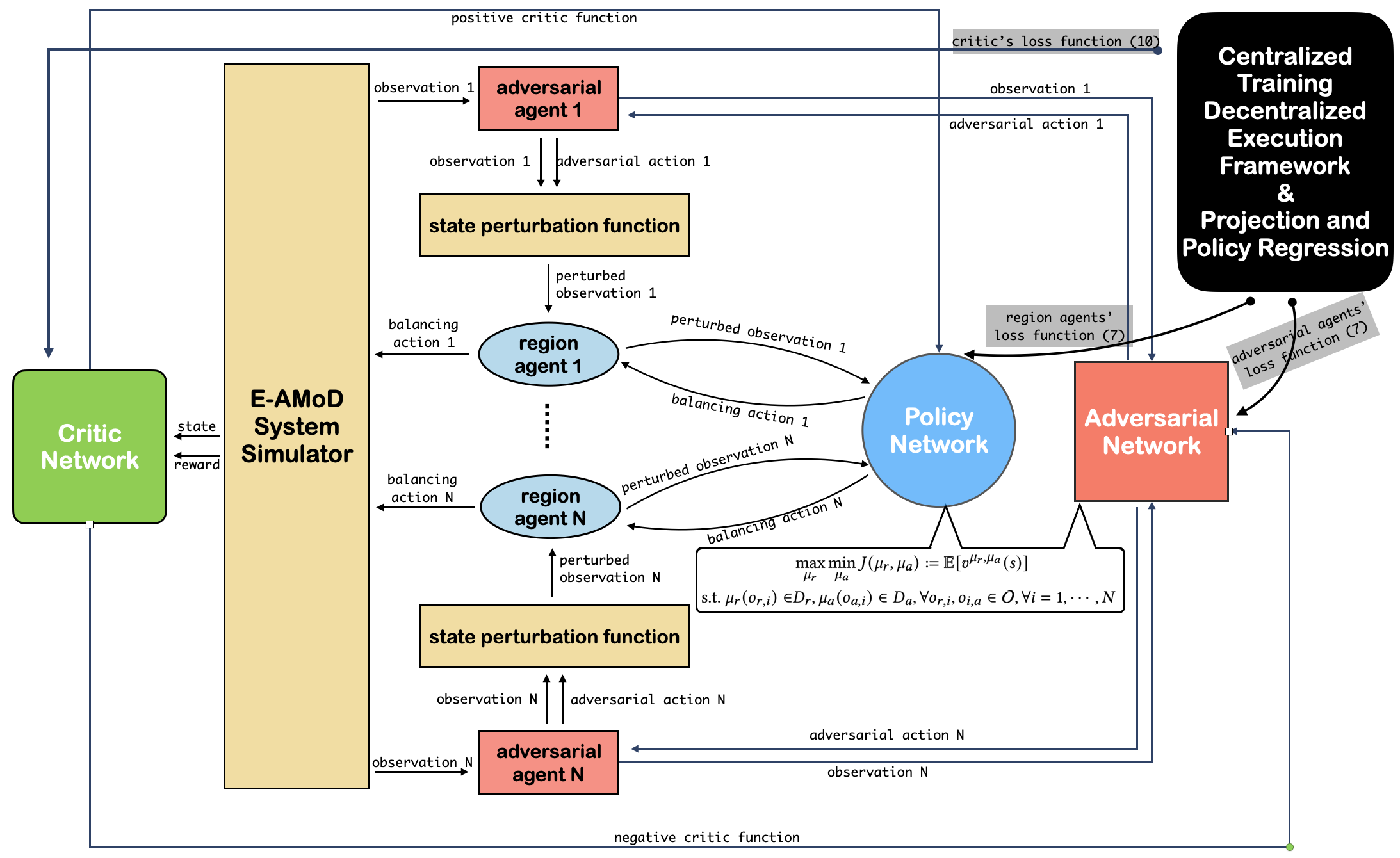}
	\vspace{-18pt}
	\caption{A brief framework structure of the proposed REBAMA algorithm.}
	\label{fig_alg}
 	\vspace{-15pt}
\end{figure}

In this section, we propose a novel MARL algorithm named \textbf{r}obust \textbf{E}-AMoD \textbf{Ba}lancing \textbf{MA}RL (REBAMA, its brief framework structure is in Fig. \ref{fig_alg}) to solve the proposed min-max problem \eqref{minmax} with policy constraints, i.e.,
\begin{align}
    \label{minmax_constraint}
    \max_{\mu_r}\min_{\mu_a}J (\mu_r, \mu_a) := \mathbb{E}[v^{\mu_r, \mu_a}(s)] \\
    \text{s.t. }   \mu_r(o_{r,i}) \in {D}_r,\ \mu_a(o_{a,i}) \in {D}_a, \nonumber\\ \forall o_{r,i}, o_{a,i} \in \mathcal{O}, \forall i = {1, \cdots, N}, \nonumber
\end{align}
where the constraints domains $D_r = \{a_{r,i} = (\overline{p}, \overline{q}) \in \Delta^{n_i} \times \Delta^{n_i} |\sum_j \bar p_j = \sum_j \bar q_j \equiv 1\}$ and ${D}_a = \{ a_{a,i} = [\delta_d,\delta_c, \delta_v] : d_l \leq \delta_d \leq d_u; c_l \leq \delta_c \leq c_u; v_l \leq \delta_v \leq v_u\}$. 
In summary, the produced rebalancing and charging actions should satisfy the normalization constraints, and adversarial agents' actions should meet the box constraints as we have defined in section \ref{sec_game}. 
In this way, ${D}_r, {D}_a$ are convex and closed sets. 

\subsection{Centralized Training Decentralized Execution Framework}
We design an actor-critic algorithm under the centralized training decentralized execution (CTDE) framework. Value-based RL algorithms such as Q-learning \cite{mnih2015human_nature} cannot solve our problem since they cannot be used for continuous spaces. 
Traditional RL algorithms such as policy gradient and actor-critic are poorly suited for multi-agent environments, because the environment becomes non-stationary from any individual agent's perspective as other agents' policies change during training. The CTDE framework is proposed to solve this non-stationary issue and has been utilized in several MARL algorithms~\cite{lowe2017multi}. It extends actor-critic methods that the critic can use extra information about the policies of other agents to ease training, while the actor is only permitted to use local information. In general, in an $N$-agent Markov game with a set of agent policies $\{ \mu_1,...,\mu_N\}$ parameterized by $\{ \theta_1,...,\theta_N\}$ respectively, the critic $q_i(s, a_1,...,a_N)$ is a centralized action-value function that computes the Q-value for agent $i$ based on all agents' action and state information. The gradient for updating policy $\mu_i$ using multi-agent deep deterministic policy gradient algorithm (MADDPG) \cite{lowe2017multi} is:
\begin{align}
\label{maddpg:gradient}
    \nabla_{\theta_i}J(\theta_1, \cdots, \theta_N) = \mathbb{E}_{\mathcal{D}} [\nabla_{\theta^i} \mu_i(o_i) \nabla_{a_i} q_i(s,a_1,...,a_N)],
\end{align}
where $a_i = \mu_i(o_i)$, $o_i$ is agent $i$'s observation, $\mathcal{D}$ is a replay buffer. 

\subsection{Projection Procedure in Policy Training}
We first design a projection procedure in policy training to satisfy the policy constraints. We define a projection operator as
$\Pi_\mathcal{A}(a) = \argmin_{z \in \mathcal{A}} \|a - z\|_2,$ 
where $\mathcal{A}$ is the action space (constrained space), $a$ is an input (potentially infeasible action), $a^p = \Pi_\mathcal{A}(a)$ is the output (projected action). When all constraints on action are linear, the constrained action space $\mathcal{A}$ is a polytope and we can use $\mathcal{H}$-representations to express the constraints as an intersection of $u$ half-spaces: $\mathcal{A} = \{a : \textbf{c}_j^{T}a \leq \textbf{e}_j, j = 1,...,u\}$. Thus, given $\mathcal{A}$ and a potentially infeasible action $a$, we can get a projected action $a^p$ by using Dykstra's projection algorithm~\cite{Gaffke1989}. 
It can generate sequences $a_j^n$ and $I^n_j$, $n \in \mathbb{N},j = 1,...,u$, which is recursively computed as follows:
$$a_j^n = \mathcal{P}_j(a_{j-1}^n - I_j^{n-1}| \mathcal{A}),
        I_j^n = a_j^n - (a^n_{j-1} - I_j^{n-1}),$$
with initial values $a^0_u = a$, $I_j^0 = \textbf{0}$, $j = 1,...,u$ and notations $a_0^n = a^{n-1}_u, n\in \mathbb{N}$, $\mathcal{P}_j$ the projection onto the $j$-th halfspace. The sequence of variables $a_j^n$ and $I_j^n$ is guaranteed to converge to the projection since $\mathcal{A}$ is the intersection of closed convex sets~\cite{Gaffke1989}.

\subsection{Robust E-AMoD Balancing Multi-Agent Reinforcement Learning Algorithm}
We summarize our proposed Robust E-AMoD Balancing Multi-Agent Reinforcement Learning Algorithm (REBAMA) in Algorithm \ref{alg: summary}. REBAMA adopts actor-critic and centralized training decentralized execution framework to deal with continuous state and action spaces, and the non-stationary issues in multi-agent environments. REBAMA also utilizes a policy regression scheme during training to accommodate the policy constraints.

In our algorithm, we use parameterized policies for region and adversarial agents, denoted by $\mu_r(\cdot | \theta^r)$ and $\mu_a(\cdot | \theta^a)$, respectively. Since region and adversarial agents get opposite rewards, their value functions are also opposite. We therefore use one parameterized critic $q(\cdot|\theta^q)$ in our algorithm so that $-q(\cdot|\theta^q)$ becomes the critic of adversarial agents. The superscripts of the parameters are occasionally omitted when there is no confusion. 

We first initialize policies and critic neural networks, target neural networks, and the global state of the environment. For each $i \in \mathcal{N}_a$, adversarial agent $i$ receives the input $o_{a,i} := \{s_i,s_{\mathcal{N}(i)}\}$ of its policy, and get its action $a_{a,i} = \mu_a (o_{a,i}|\theta^a)$. Then according to the perturbation rule defined in \eqref{def:map}, we get the corresponding region agent $i$'s input $o_{r,i} :=\{ \Tilde{s}_i, s_{\mathcal{N}(i)}\}$, which contains perturbed state information. 
The balancing decision $a_{r,i} = \mu_r (o_{r,i}|\theta^r)$ is then calculated. We get the adversarial and region agents' joint actions $a_a, a_r$ by repeating the above steps for all $i$. After executing the region agents' joint actions, we store a transition $(s, {a}_{r}, {a}_{a}, r, s^{\prime})$ in a replay buffer $\mathcal{D}$, where $r$ is a shared reward, and $s^{\prime}$ is the next state. 

At each training iteration, we sample a minibatch of transitions $\mathcal{B}$ from $\mathcal{D}$ and update policies using the following loss function with policy regression:
\begin{align}
\label{policy_reg}
    &\mathcal{L}^p(\theta) = \sum_{\mathcal{B}} \|\mu(o|\theta) - \hat{a}\|^2,\\
\label{def:a}
    &\hat{a} = (1-\delta)\Pi_\mathcal{A}(\mu(o|\bar{\theta})) + \delta x,\\
\label{def:x}
    &x = \argmax_{x \in \mathcal{A}} x^{T}\nabla_a \phi q(s, a_a^\prime, a_r^\prime|\bar\theta^q),
\end{align}
where $\delta$ is a positive step size, $a_a^\prime = \Pi_{\mathcal{A}_a}(\boldsymbol{\mu}_a(\boldsymbol{o}_a | \bar\theta^a))$, and $a_r^\prime = \Pi_{\mathcal{A}_r}(\boldsymbol{\mu}_r(\boldsymbol{o}_r| \bar\theta^r))$. $\phi$ is a coefficient whose value is given later. $\bar\theta^a, \bar\theta^r, \bar\theta^q$ are snapshots of the current actor and critic parameters. We use $\boldsymbol{\mu}_a(\boldsymbol{o}_a) := (\mu_a(o_{a,1}), ... ,\mu_a(o_{a,N}) )$ to denote the joint adversarial policy and $\boldsymbol{\mu}_r(\boldsymbol{o}_r)$ has a similar definition. When updating region agent's policy i.e. $\theta_t = \theta^r_t$, we have $o = o_{r,i}, \phi = 1, a = a_{r,i}^\prime, \mathcal{A} = \mathcal{A}_{r,i}$. Otherwise $o = o_{a,i}, \phi = -1, a = a_{a,i}^\prime, \mathcal{A} = \mathcal{A}_{a,i}$. $i$ is the region index. The centralized action-value function $q$ is updated using the following loss function:
\begin{align}
\begin{split}
    &\mathcal{L}(\theta^q) = \sum_{\mathcal{B}} [q(s, a_a, a_r) - y ]^2,\\
    &y = r + \gamma q^{\prime}(s^{\prime}, a_a^{\prime}, a_r^{\prime})|_{a_a^{\prime}=\boldsymbol{\mu}_a(\boldsymbol{o}_a), a_r^{\prime} = \boldsymbol{\mu}_a(\boldsymbol{o}_r)},
    \end{split}\label{update:critic}
\end{align}
where $q^\prime$ is the target critic neural network.
\begin{algorithm}
 \caption{REBAMA Algorithm}
 \label{alg: summary}
 \begin{algorithmic}
 \renewcommand{\algorithmicensure}{}
 \STATE \textbf{Initialize} critic $q(s,a_a,a_r)$, region policy $\mu_r$ and adversarial policy  $\mu_a$, parameterized by $\theta^q,\theta^r,\theta^a$ respectively.
 \STATE \textbf{Initialize} replay buffer $\mathcal{D}$, target networks $q^{\prime}, \mu_r^{\prime}, \mu_a^{\prime}$ with weights $\theta^{q\prime} \leftarrow \theta^q, \theta^{r \prime} \leftarrow \theta^r, \theta^{a \prime} \leftarrow \theta^a$.
  \FOR {$\text{episode} = 1 \text{ to } M$}
  \STATE \textbf{Save} weights $\bar{\theta}^r \leftarrow \theta^r, \bar{\theta}^a \leftarrow \theta^a, \bar{\theta}^q \leftarrow \theta^q$,
  \FOR{ $t = 1 \text{ to } T$}
  \STATE \textbf{Receive} the initial state $s_1$,
  \STATE \textbf{Adversarial} agents select actions $a_{a,t,i} = {\mu}_a(o_{a,t,i}|\bar{\theta}^a)$ and region agents select actions $a_{r,t,i} = {\mu}_r(o_{r,t,i}|\bar{\theta}^r)$ where $i = 1,...,N$.
  \STATE \textbf{Execute} joint action ${a}_{r,t}$, then get reward $r_t$, next state $s_{t+1}$ and store $(s_t, {a}_{r,t}, {a}_{a,t}, r_t, s_{t+1})$ in $\mathcal{D}$.
  \STATE \textbf{Sample} a random minibatch $\mathcal{B}$ from $\mathcal{D}$.
  \STATE \textbf{Update} policies $\mu_r, \mu_a$ using loss function~\eqref{policy_reg}.
  \STATE \textbf{Update} critic $Q$ according to loss function (\ref{update:critic}).
  \STATE \textbf{Update} target networks $\theta^{q\prime} \leftarrow \tau\theta^q + (1-\tau)\theta^{q\prime}$, $\theta^{r\prime} \leftarrow \tau\theta^r + (1-\tau)\theta^{r\prime}$, $\theta^{a\prime} \leftarrow \tau\theta^a + (1-\tau)\theta^{a\prime}$.
  \ENDFOR
  \ENDFOR
 \end{algorithmic} 
 \end{algorithm}
\begin{Proposition}
When there are no constraints on the action space, the policy gradient of REBAMA can be directly calculated by $\phi \sum_{\mathcal{B}} \nabla_{a} q(s,a_a,a_r | \bar\theta^q)\nabla_{\theta}\mu(o|\theta)$ where $a_a = \boldsymbol\mu_a(\boldsymbol{o}_a|\bar\theta^a), a_r = \boldsymbol\mu_r(\boldsymbol{o}_r|\bar\theta^r)$. 
\end{Proposition}
\begin{proof}
When there are no constraints on the action space, $\hat{a}$ is equivalent to $\mu(o|\bar\theta) + \eta \nabla_{a} \phi q(s,a_a^\prime,a_r^\prime|\bar\theta^q)$ where $\eta$ is a positive step size, $a_a^\prime = \Pi_{\mathcal{A}_a}(\boldsymbol{\mu}_a(\boldsymbol{o}_a | \bar\theta^a))$, $a_r^\prime = \Pi_{\mathcal{A}_r}(\boldsymbol{\mu}_r(\boldsymbol{o}_r| \bar\theta^r))$ Then the gradient of \eqref{policy_reg} equals to:
\begin{align*}
    \nabla_{\theta} \mathcal{L}^p =  2\eta\phi \sum_{\mathcal{B}} \nabla_{a} q(s,a_a^\prime,a_r^\prime|\bar\theta^q)\nabla_{\theta}\mu(o|\theta),
\end{align*}
Comparing to the deterministic policy gradient in~\eqref{maddpg:gradient}, we can see that the policy updating procedure in REBAMA with a learning rate of $\eta_1$ is equivalent to that in deterministic policy gradient with a learning rate of $2|\mathcal{B}|\phi\eta\eta_1$.
\end{proof}

%% file: sec_005_exp.tex
We use real-world E-taxi data from Shenzhen city to conduct experiments. Three different data sets~\cite{dro_he, wang2019shared} including E-taxi GPS data (vehicle ID, locations, time and speed, etc), transaction data (vehicle ID, pick-up and drop-off time, pick-up and drop-off location, travel distance, etc) and charging station data (locations, the number of charging points, etc) are used to build an E-AMoD system simulator as the training and evaluation environment. 
To test the robustness of our proposed robust method, we inject a Gaussian noise follows $\mathcal{N}(0,1)$ into the state when testing vehicle balancing methods. The simulated map is set as a grid world. The policy networks and critic network are two-layer fully-connected networks, both with 30 nodes. For the policy networks and critic network, the first and second hidden layers' activations are Tanh and ReLu, the output layer is Softmax and Linear,   respectively. We set the batch size $= 600$, the discount factor $= 0.99$, the time interval $= 0.5$ hour, and use the Adam optimizer with a learning rate of $0.001$. For fair comparisons, we use fixed random seeds to initialize simulations in the testing.

\subsection{Comparison of Robust MARL Method and Robust Optimization Method}
We first compare our robust MARL method (REBAMA) with robust optimization method~\cite{dro_he} in table~\ref{tab:model-base}. We test these two methods in the simulator for 5 times with 5 different fixed random seeds when state uncertainty is present,  then compare various average metrics. The proposed REBAMA algorithm performs better on average reward, average fairness of supply-demand ratio and average fairness of charging utilization which is increased by 8.21\%, 9.42\%, 8.29\%, respectively. Though the robust optimization method also considers demand and supply uncertainties, it heavily depends on predetermined parameters such as probability transition function, prediction models, etc. This heavy dependence makes the robust optimization method can be only robust for specific scenarios and not sufficient to capture the complexity of E-AMoD systems. Our REBAMA algorithm is better at capturing E-AMoD system’s demand and supply uncertainties.
\begin{table}[]
\centering
\vspace*{6pt}
\caption{Robust MARL \textbf{VS} Robust Optimization Method}
\vspace*{-6pt}
\begin{tabular}{cccc}
\hline
$\text{Metric}$     & \text{Robust Opt}     & \text{Robust MARL} & \text{Increasing Rate} \\ \hline
$\text{average reward}$ & $-15.83$ & $-14.53$  & $\uparrow8.21\%$  \\\hline
$\text{average } u_s$ &  $-7.01$ & $-6.35$  &  $\uparrow9.42\%$ \\\hline
$\text{average } u_c$ &  $-8.82$ & $-8.18$ &  $\uparrow8.29\%$ \\\hline
\end{tabular}
\label{tab:model-base}
\vspace{-10pt}
\end{table}
\begin{figure}[ht]
	\centering
	\includegraphics [width=0.48\textwidth]{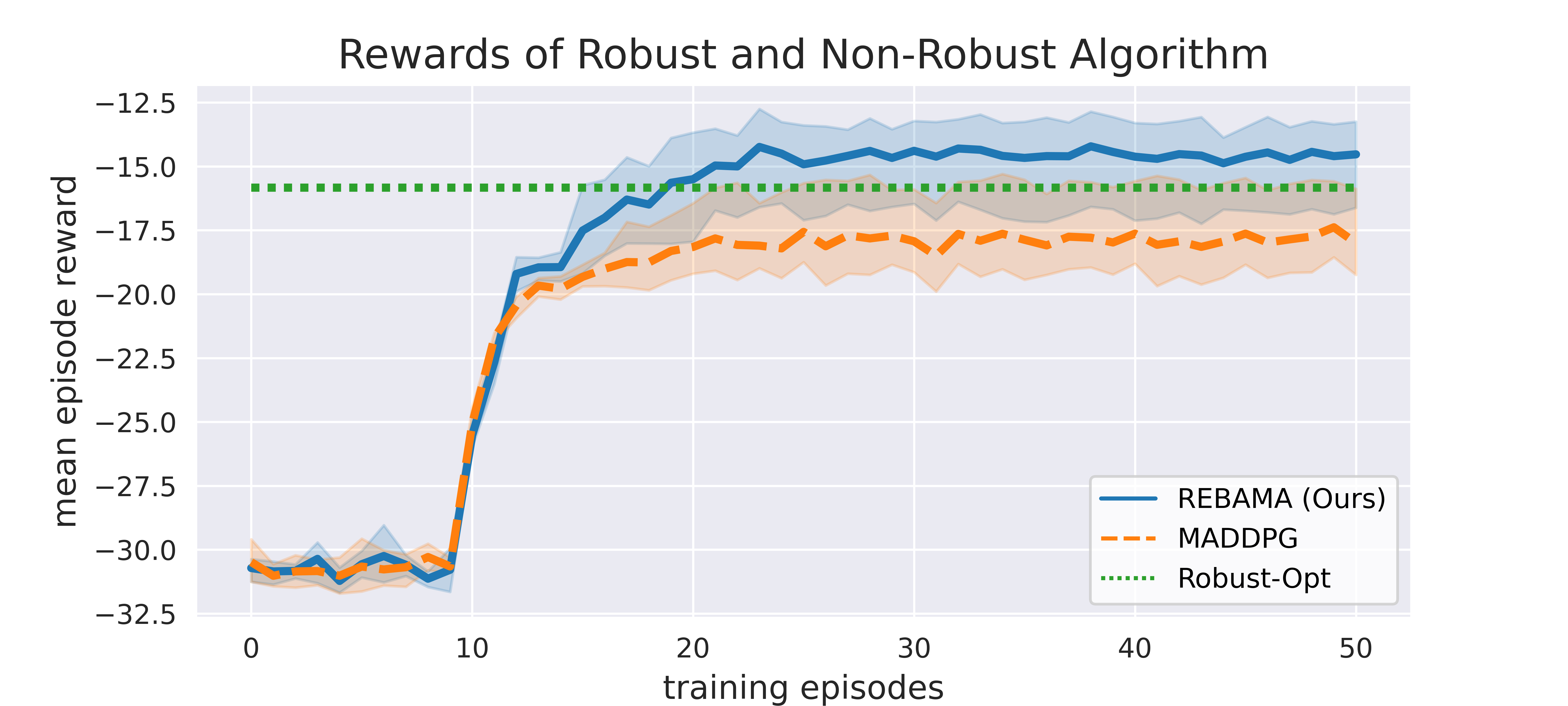}
 	\vspace{-18pt}
	\caption{Compared to MADDPG, a non-robust MARL algorithm, by using our REBAMA method, the mean episode reward is increased by 19.28\% when state uncertainty is present.}
	\label{fig:reward}
	\vspace{-10pt}
\end{figure}
\begin{figure}[ht]
	\centering
	\includegraphics [width=0.48\textwidth]{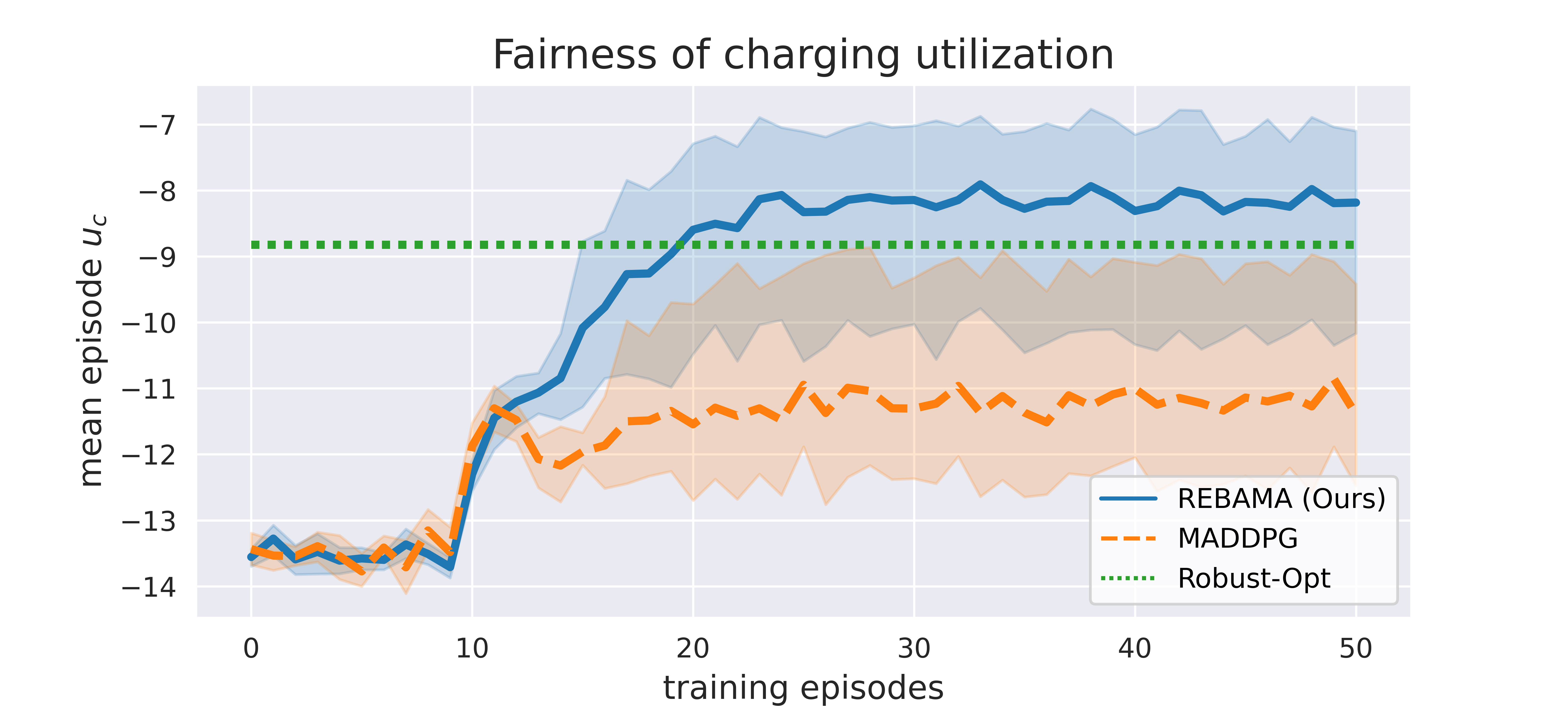}
	\vspace{-18pt}
	\caption{Compared to MADDPG, a non-robust MARL algorithm, by using our REBAMA method, mean episode charging fairness is increased by 28.18\% when state uncertainty is present.}
	\label{fig:u}
	\vspace{-10pt}
\end{figure}
\begin{figure}[ht]
	\centering
	\includegraphics [width=0.48\textwidth]{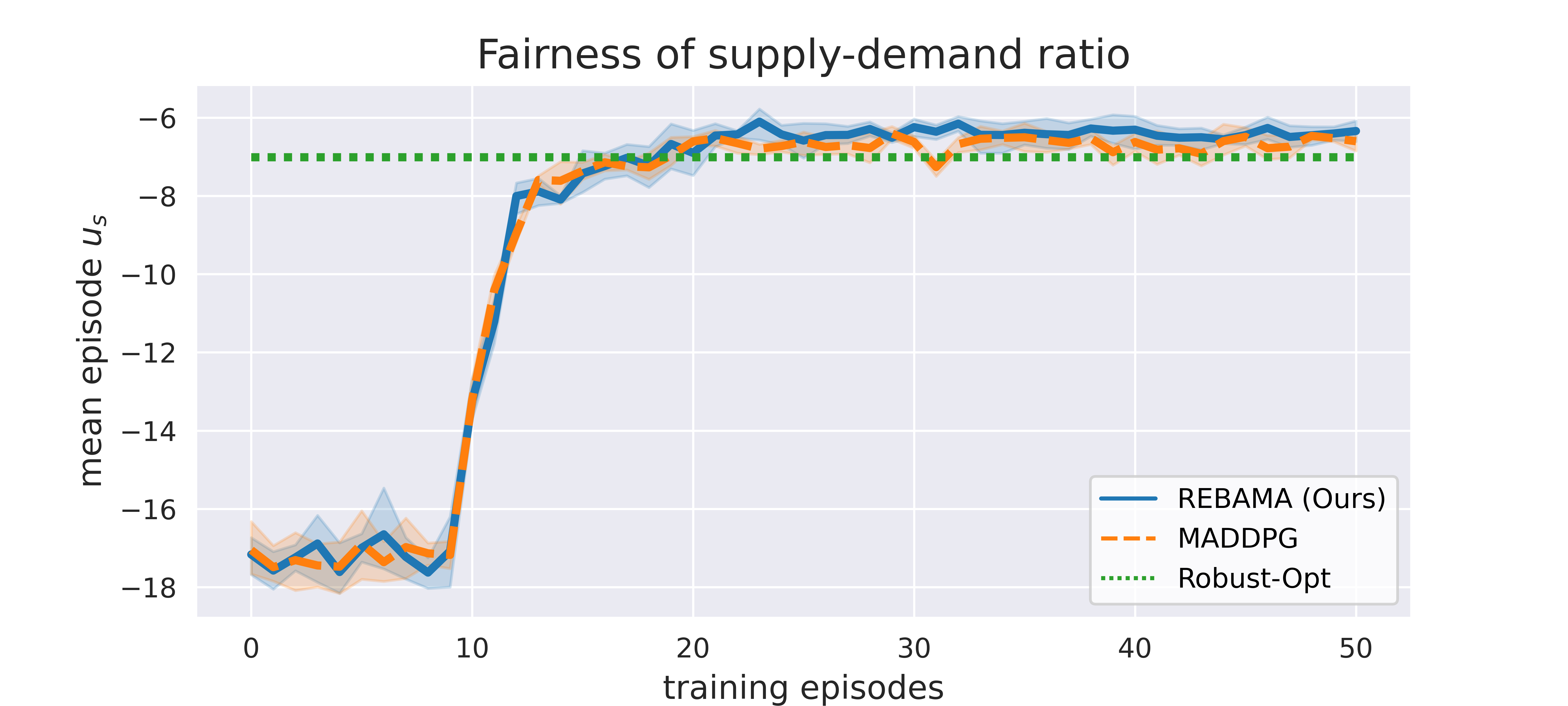}
 	\vspace{-18pt}
	\caption{Compared to MADDPG, a non-robust MARL algorithm, by using our REBAMA method, mean episode supply-demand fairness  is increased by 3.97\% when state uncertainty is present.}
	\label{fig:sr}
 	\vspace{-10pt}
\end{figure}

\subsection{Comparison of Robust and Non-Robust MARL Methods}
We then compare our proposed robust multi-agent reinforcement learning (MARL) approach to a commonly-used and well-performed non-robust MARL algorithm, i.e. multi-agent deep deterministic policy gradient (MADDPG)~\cite{lowe2017multi} which does not consider supply or demand uncertainties. We train our models until convergence and then evaluate them by averaging various metrics for 5 iterations. To avoid constraint violations, we also apply policy regression to MADDPG.
We report the mean episode rewards as a function of the training episodes in Fig. \ref{fig:reward}. The proposed REBAMA algorithm learns a better policy in terms of mean episode reward which is increased by 19.28\% compared with MADDPG when state uncertainty is present. Higher rewards mean the city has at least a better balanced supply-demand ratio or a fairer charging utilization. 

Since a reward is a weighted sum of fairness of charging utilization $u_c$ and supply-demand $u_s$ defined in \eqref{def:reward} and \eqref{def:u}, we also compare mean episode $u_c$ and $u_s$ of each episodes in Fig.~\ref{fig:u} and Fig.~\ref{fig:sr}. A larger $u_c$ or $u_s$ means a better balanced or fairer service across the city. Our REBAMA algorithm learns a better policy compared to the non-robust algorithm, where the charging utilization fairness and supply-demand fairness are increased by 28.18\%, 3.97\% respectively. Furthermore, we find that REBAMA improves the charging utilization greatly compared to the non-robust algorithm. This can be explained by the fact that charging utilization is very sensitive to EAVs’ supply-side uncertainty.

\subsection{Effectiveness of Policy Regression and Projection}
\begin{figure}[!t]
	\centering
	\includegraphics [width=0.48\textwidth]{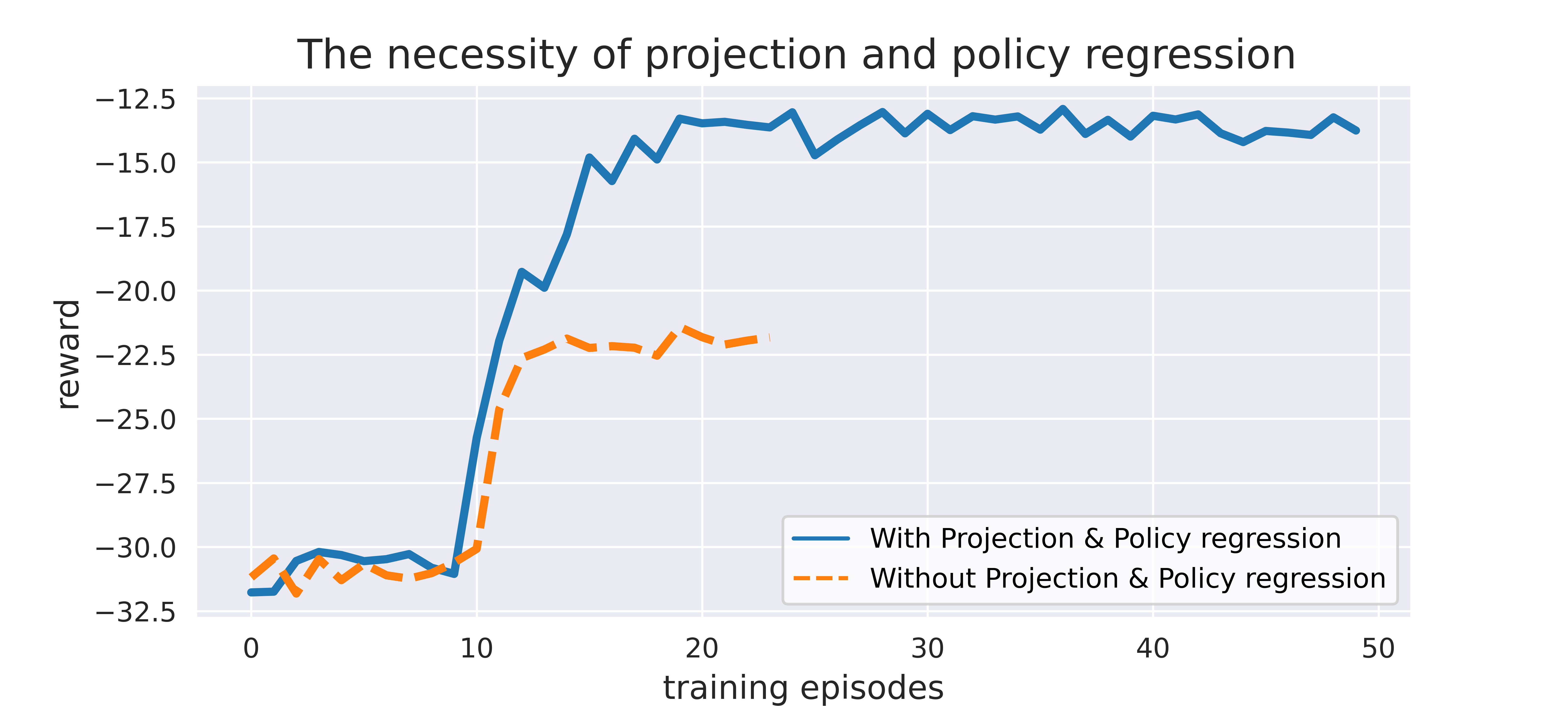}
	\vspace{-18pt}
	\caption{Projection and policy regression procedure in the training process is effective to ensure feasible actions and training's continuity.}
	\label{fig:projection}
	\vspace{-15pt}
\end{figure}

In Fig.~\ref{fig:projection}, we compare our REBAMA and the MADDPG algorithm without policy regression and projection to study the effectiveness of adding projection and policy regression steps in the training process. Without them, MADDPG stops training in earlier episodes when the policy outputs an infeasible action. REBAMA can handle this interruption problem then the training process will not stop before the training is done.

